# Germanium Metasurface for MWIR Polarization-Sensitive Stokes Thermal Imaging at 4-micron wavelength.

**Hosna Sultana**

*School of Electrical & Computer Engineering, University of Oklahoma, 110 W Boyd St, Norman, OK 73019, USA*

*hosna@ou.edu*

**Abstract:** The mid-wave (MWIR) spectral range can provide a larger bandwidth for optical sensing and communication when the near-infrared band gets congested. As optical sensing becomes a robust technique for digital imaging and object recognition, this range of thermal imaging needs to convey more information, which can be unraveled from polarization-sensitive detection by integrating the metasurface of the subwavelength scale structured interface to control light-matter interaction. To enforce metasurface-enable simultaneous detection and parallel analysis of polarization states in a compact footprint for 4-micron wavelength, we design a high-contrast Germanium metasurface with an axially asymmetric triangular nanoantenna of the thickness of 0.525 times the wavelength. First, we optimized linear polarization separation of a 52-degree angle with 50% transmission efficiency, holding the meta-element aspect ratio from the 3.5-1.67 range. The transmission modulation in terms of periodicity and lattice resonance for the phase gradient high contrast dielectric grating in correlation with scattering cross-section for both 1D and 2D cases has been discussed, which enables the reduction of the aspect ratio to meet the nanofabrication challenge. Further, by employing the geometric phase, we get 40% and 60% transmission contrast for linear and circular polarization states, respectively, and reconstruct the Stokes vectors and output polarization states. Without spatial multiplexing, this single metasurface unit cell can perform well for the division of focal plane Stoke thermal imaging with an almost 5-degree tilted incidence, excellent refractive index, and height tolerance.

## 1. Introduction:

Polarization is the inherent property of light, in which the information remains encoded in the Poincaré space and can be decoded knowing the state of polarization (SOP) of the light. When light is transmitted or reflected by any structure, the propagating SOP with respect to the incident SOP can provide a lot of information if we can detect it, which is a rich field of polarimetry. Polarization states of light give the hidden features of an object, like surface pressure and orientation, multiple layers, and transparency, to ensure better object recognition. Three main detection methods exist: division of amplitude, division of aperture, and division of focal planes. There are 8-pixel to 4-pixel and even 3-pixel techniques for maneuvering polarization filtering FPA technology operates without metasurface [1,2]. Polarization-sensitive imaging technologies have been around for several decades but are still limited by mechanical switching and bulky optics. Previously, the division of aperture was utilized for MWIR polarimetric imaging [3]. However, metasurface-enabled on-chip polarimetry is the technology trend now aimed at compact imaging devices.

 The increasing application for computational image analysis, pattern recognition, machine vision, Lidar, and remote sensing largely depended on polarization selective imaging for getting edges and corners for better image contrast. Various technologies have been applied to MWIR polarization image sensors, such as high-speed rotating polarization analyzers, where the image of different polarized angles is captured by swiftly rotating the analyzer. For a scenario of a rapidly changing object's features or target's speed, we need to obtain the Stokes parameters simultaneously without time delay, so the recent technique of integrating compact optical metasurface for division of focal plane polarization-sensitive imaging needs further implementation for MWIR thermal sensing. Though the infrared polarized imaging technique is utilized for various commercial and defense applications, it is still very expensive.

In recent years, the subwavelength scale artificially structured surface, called metasurface, has been proven a reliable technology to unravel this phenomenon in an efficient way [4–11]. Dielectric metasurface diffraction grating structure has been utilized for polarimetry, polarization converters, holography, optical sensing, and many other applications. Optimized metasurface devise structures are already in commercial application in the visible and near-infrared range; very few pieces of literature reported the metasurface design for polarimetric application in the MWIR range, which needs more attention [12–17]. Excellent work presented by Chen et al. for simultaneous wavelength multiplexing polarimetry, which behaves as a cascade of parallel filter, polarizer, and waveplate for the 3-4.5 μm range, overcoming many challenges. However, the limitation of the design is the detection of only one set of orthogonal polarization states at a particular wavelength and with the nanoantenna of a high aspect ratio ~ 30 [17]. Yan et al. proposed the division of a focal plane metasurface imaging mask with silicon metasurface using a polarization-controlled phase modulation technique. However, it is for Long-Wave infrared 10.6 μm wavelength [16]. Yin et al. show the Si over SiO2 substrate has better transmission efficiency but is operational in the near-infrared [18]. The phase change material of $Ge_2Sb_2Te_5$ (GST) allows more tunability owing to its inherited amorphous to the crystalline transition property with polarization switching, but it is useful for LWIR [19,20]. To investigate the polarization-controlled and achromatically on-axis focused optical vortex beams for 3.5 to 5 μm, Ou et al. came up with all Si metasurface optimization [20]. They show promising functionality, even being capable of switching the topologically charged states. Yet the high aspect ratio and only two orthogonal polarization-controlled devices pose limitations for on-chip polarimetry.

The variety of metasurfaces consists of staking geometry, and layered structures are evolving, which poses a challenge to nanofabrication. With a vertically stacked photodiode, the nanowire polarizer filters have been reported [21], and Hybrid image sensors for detecting angular light, intensity, and polarization [1] but very complicated fabrication steps are required, which can be done with single-step lithography. We propose a simple Ge-metasurface for simple step fabrication for stoke imaging optimized for 4 μm wavelength, which is the high transparent atmospheric window.

## 2. Design principle and simulation

**2.1 Material library, design parameter, and simulation methodology:**

The present work uses the axially asymmetric triangular meta-element of Germanium of refractive index $n_{Ge}$= 4.025 at 4 μm wavelength for the metasurface design with the finite difference time domain (FDTD) simulation package from Ansys-Lumerical Inc. The low refractive index substrate $CaF_2$ has been used for the transmission mode design, which is suitable for this wavelength and for integrating the metasurface with an uncooled PbSe focal plane array (FPA). First, gradually increasing the height, we check the full phase coverage for a periodicity of 2500 nm. Then, for a fixed height that provides full phase coverage, we varied the length and base of the triangle nanoantenna, as shown in Fig. 1, the phase and transmission library for the single meta-element. A periodic boundary has been implemented on the sides and PML boundary to the light propagating direction from the substrate towards the nanoantenna, placing the DFT monitor several wavelengths away. For the scattering cross-section, we use all PML boundaries with a total-field scattered-field (TFSF) source. Due to the high refractive index of Ge, the lateral dimension is reduced to $n_{Ge}.L/\lambda_0 = 0.6$ to 1.4 to find a homogeneous region of phase and transmission evolution.

Unlike symmetric rectangular meta-elements, the triangular shape accumulates somewhat different $\varphi_x$ and $\varphi_y$ with increasing height. It takes $\geq 2\lambda$ propagation length inside the meta-element for sufficient propagation phase control. For the same height, lateral dimension, and periodicity, the rectangular meta-element can give more phase control, but the triangular meta-element gives more transmission [22]. The $\varphi$, T library is presented in Fig.1, which shows some phase discontinuity with anti-crossing transmission behaviors for some spatial dimensions. For this type of high-contrast dielectric structure, resonant absorption decreases the transmission, but it mitigates the resonance loss for the triangular geometry. Within this element dimension, the $2\pi$ phase control is difficult with the height reduction of the nanoantenna, as indicated in Fig 2(c). So, Fig. 2(a) shows selecting the appropriate elements for $2\pi$ propagation phase gradient. A 2150 nm height 1-dimensional (1D) metasurface gives a maximum of about 60% transmission to the prominent order, whereas a 2750 nm height design can give 78%. For the practical concern of nanofabrication, we aim to reduce the aspect ratio more; we went down to a height of 2150 nm because increasing the lateral dimensions is limited by periodicity.

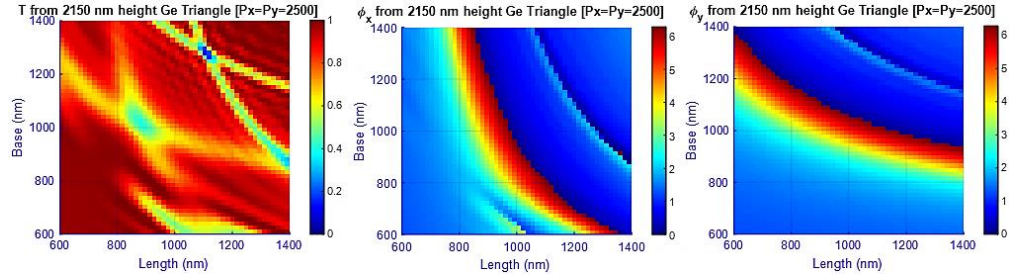

Fig. 1. Transmission and phase library for 2150 nm Ge triangular nanoantenna.

**2.2 Lattice resonance to enhance transmission:**

To overcome the transmission loss with height reduction, we implement the strategy of interacting nano-antenna mentioned elsewhere [22,23]. For the periodicity range of the mutually interacting nanoantenna when $(n_{substrate} \times periodicity)/wavelength < 1$ holds [22–24]. In this strategy to utilize the lattice resonant for boost transmission, we reduce the periodicity, and the comparison of the same unit cell element has been provided in Fig.2 (a), (b) where we construct the 7x1 element 1D metasurface.

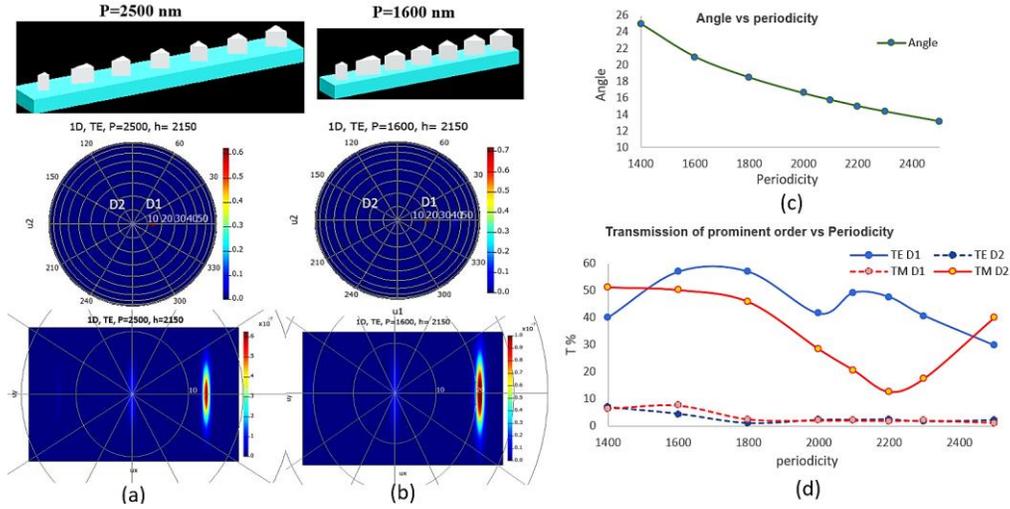

Fig. 2. (a) 1D metasurface design for P=2500 nm and the efficiency at the detector position. (b) Same for P=1600 nm. (c) The periodicity Vs. beam deflection angle at the prominent order. (d) The effect on the transmission of varying the periodicity for the 2D LP separation design.

From the linear polarization (LP) separation design, we see in Figure 2(d) that the transmission efficiency is highest around the periodicity Px=Py=1600 nm for both prominent transmission orders at the two LP separation order positions, D1 and D2. Here by $P_x=P_y$ we mean periodicity of the single element unit cell. So, for the normalized lattice constant $P_x/\lambda_0$ =$P_y/\lambda_0$ = 0.625 to 0.35 ratios, the best transmission comes when $P_x=P_y = 0.4 \lambda_0$. However, we can say that the propagation phase correlation between the meta-elements in this range remains intact. Although the generalized Snell's law explains the phase gradient metasurface, which has served the purpose for decades, its effect on the scattering cross-section is not discussed much. In Fig. 3, Here, we investigate the scattering cross-section for the phase gradient metasurface grating (PGMG) for both TE and TM incidence and compare the total transmission efficiency for the 2D metasurface.

The scattering cross-section from 1D metasurface when the element of the same height is arranged in a $2\pi$ propagation phase gradient (PPG) along the x-axis called PGMG (magenta line) and when the element of the same height is placed at constant $\pi$ phase along x-axis called constant phase (CPMG) (cyan line) with varying periodicity for TM incidence presented in Fig. 3 (a). Though around $P_x/\lambda_0 = P_y/\lambda_0 = 0.625$, both scattering cross-sections coincide, but with decreasing the $P_x=P_y$, the scattering cross-section increases for both the PGMG and CPMG when the normal incident light has the E-field parallel to the direction of the PPG. Still, for the 1D metasurface, the trend of the difference in scattering cross-section ($\Delta\sigma_{SC} = \sigma_{SC, PGMG} - \sigma_{sc, CPMG}$) has a linear correlation with periodicity.

As we look at the 2D-PGPM of Figure 3(b), we see that with decreasing the $P_x=P_y$, the scattering cross-section decreases with a gradual increase around $0.5\lambda_0$-$0.55\lambda_0$ for the PGMG when the normal incident light has the E-field parallel to the direction of the PPG. Here, we are only interested in the TM or E-field parallel to the direction of the PPG case because, in the 2D metasurface design, the PPG has been assigned oppositely towards the x-axis for x and y-polarized light. So, the TM incidence is affected the most due to the surface lattice resonances (SLR). We can see in Fig. 3(c) that with the periodicity decrease, the total transmission increases, and reflection decreases for TE incidence, but there is no smooth relation for TM incidence. For the negligible absorption of Ge, the scattering conveys the extinction. From the Kerker condition, we know the forward or backward scattering can be tuned by in-phase and out-of-phase electric dipole (ED) and magnetic (MD) moments, respectively, among this type of subwavelength meta-elements [25,26]. The loss in the total transmission around $0.5\lambda_0$ - $0.55\lambda_0$ periodicity Fig. 3 (c) can be explained by the increased scattering for TM incidence (Fig. 3(b)). Additionally, T+R <1, for the $P_x=P_y = 0.35\lambda_0$ indicates the lower limit of the dipole approximation has been crossed when $P_x=P_y <$ 1500 nm [27].

Looking at the spectral response of the T from the 2D metasurface in Fig. 3 (d), we see the TM incidence suffer the T loss above the Rayleigh anomaly indicated by the black dashed line. For a single meta-element, the $\sigma_{sc}$ remains almost constant with the change of the periodicity. So, the variation of $\sigma_{sc,\ CPMG}$ for even a simple seven same meta-element arrangement in Fig. 3 (a) indicates the presence of lattice resonance. Different size of the meta-element used for creating PGMG decreases $\sigma_{sc,\ PGMG}$ because the different radiative coupling strengths may add up destructively. The ability of the high-contrast dielectric meta-element to generate both ED and MD resonance is an advantage over the plasmonic one [28,29]. For example, of the single particle under TM incidence (E along x-axes), the enhancement of E-field (right side) and H-field (left side), and the radiation pattern from smallest size (bottom row) and biggest size (upper row) is shown in Fig. 3(e). For the same periodicity, in the case of the bigger-size triangular meta-element, the E-field resonates mostly within itself. It enhances the H-field more than the smaller size one. In summary, we can say that for the 2D-PGMG, decreasing periodicity decreases the scattering and tunes the ED and MD resonance in phase more to overcome the reflection (Fig 3 (b), (c)) and increases the transmission to the prominent order more (Fig. 1(d)).

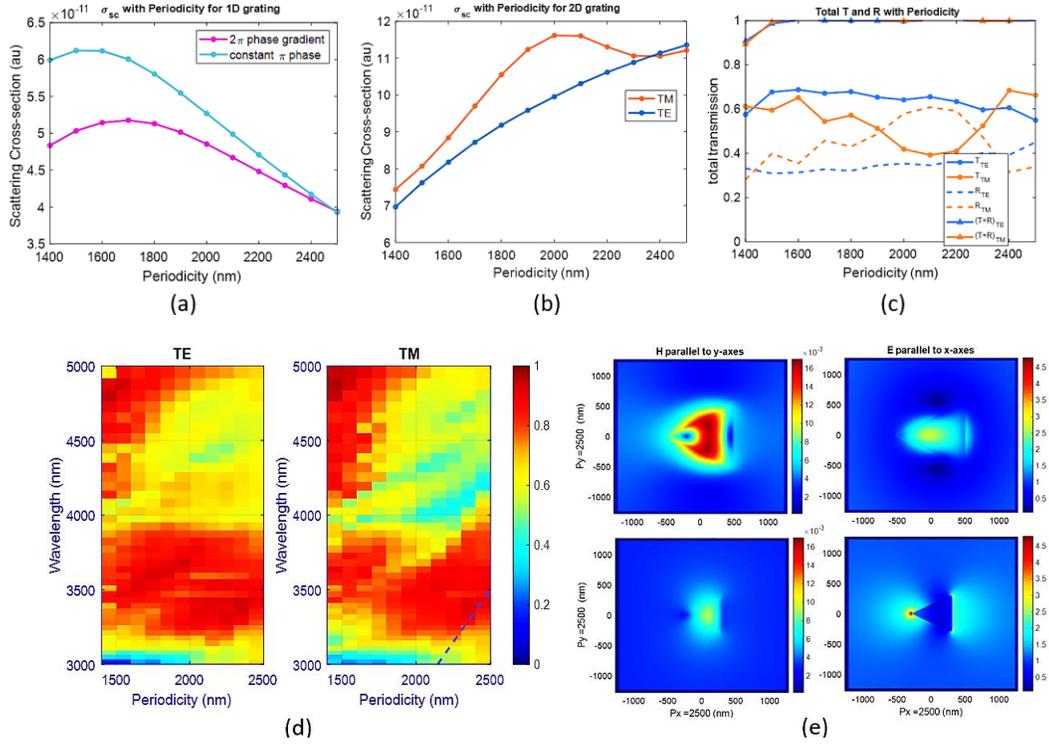

Fig. 3. (a) Scattering cross-section from 1D grating with and without phase gradient with periodicity at 4µm wavelength for TM incidence. (b) Scattering cross-section from 2D grating at 4µm wavelength with periodicity. (c) Total T and R from the same 2D grating with periodicity. (d) Spectral response of the total T from the same 2D grating from 3µm to 5µm wavelength with periodicity. (e) Difference in the E and H field enhancement and the radiation pattern from smallest size (bottom row) and biggest size (upper row) Ge-triangle nanoantenna of 2150 nm height for the same periodicity.

As our interest is in a fixed wavelength, we are not focusing on the wavelength shift of ED and MD resonances due to meta-element size and periodicity. Also, due to the variety of the size parameters and asymmetry of the geometry of our triangular meta-element, the sharp ED and MD resonant pattern is not visible due to different types of coupling strength at different lattice points [29]. However, choosing a periodicity smaller than 2000 nm for assigning the PPG at 4 μm wavelength, we avoid the particular type of lattice resonance, which has a wavelength longer than the period [29,30] scenario, where power goes to the guided mode or perpendicular evanescent order, which will remain for future study.

In this step, we decide the periodicity will be 1600 nm. So, from the same element of and T and φ of periodicity = 1.16h library, we move to a new design with periodicity = 0.74h, which increases the transmission from 40% to 50.2% for TM (x-polarized) incident and from 29.8% to 57% for TE (y-polarized) incidence as shown in Fig. 2(d). Another benefit follows, as we see in Fig. 2(c), this increase of angular separation of the deflected beam from 13.21° to 21°, which follows the PPG metasurface rule of phase profile φ and beam deflection angle θ = arcsin[φ/(2πP$_x$/λ)] as controlled by generalized Snell's law of anomalous deflection [31,32].

### 2.3 Circular Polarization (CP) control with geometric phase:

Although we found the subwavelength scale axially asymmetric nanostructure does not affect the propagation phase because of lacking rotational symmetry, it hindered the geometric phase, which is supposed to follow the Pancharatnam-Berry (PB) phase rule [33,34]. The PB phase implementation of the 1D structure has been done as convolution with a linear increment of phase. Ideally, the PB phase gives a 2θ addition in phase for a θ rotation. The transmission modulation of the incident circular polarized light can be done with the Jones matrix by the operation of the rotation matrix R(θ) as presented in equation (1) [4,35,36]:

$$t(\theta) = R(-\theta)\begin{pmatrix} t_x & 0 \\ 0 & t_y \end{pmatrix} R(\theta) \qquad (1)$$

$$|L\rangle = \frac{1}{\sqrt{2}}\begin{pmatrix} 1 \\ -i \end{pmatrix} \quad and \quad |R\rangle = \frac{1}{\sqrt{2}}\begin{pmatrix} 1 \\ i \end{pmatrix} \qquad (2)$$

Where $t_x$ and $t_y$ are the complex transmittance of light, while the incident light is polarized towards the x and y axes, respectively, when the meta-element has zero rotation. The $|L\rangle$ and $|R\rangle$ are the Jones vector of the left and right circular polarized light. The electric field of the left and right circularly polarized light is given by multiplying with the Jones vector of LCP and RCP :

$$E_L = t(\theta).|L\rangle(\theta) = \frac{(t_x + t_y)}{2}\begin{pmatrix} 1 \\ i \end{pmatrix} + \frac{(t_x - t_y)}{2} e^{i2\theta}\begin{pmatrix} 1 \\ -i \end{pmatrix} \qquad (3)$$

$$E_R = t(\theta).|R\rangle(\theta) = \frac{(t_x + t_y)}{2}\begin{pmatrix} 1 \\ -i \end{pmatrix} + \frac{(t_x - t_y)}{2} e^{-i2\theta}\begin{pmatrix} 1 \\ i \end{pmatrix} \qquad (4)$$

In equations 3 and 4, the 1st term conveys the transmitted wave with the same helicity as the incident, and the 2nd term conveys the opposite helicity with a $\pm i2\theta$ phase addition. This general PB phase rule indicates if the metasurface element cannot provide a phase of at least π/4 phase retardation, then the element will not be able to control the total phase range [37]. When we reduce the periodicity, the length/periodicity ratio increases. Consequently, the phase gets locked for specific angular rotations of the triangular meta-element, as we can see in [38]. We need to increase the $P_y$ to reduce $L/P_y$ from 0.79 to 0.63. So, we get smooth phase change by increasing only $P_y$ from 1600 nm to 2000 nm, keeping $P_x$ fixed at 1600 nm since we employed the PB phase towards y-axes only. Hence, in our final metasurface LP+CP separation design, as presented in Fig. 4 (a), we implement $P_x=1600$ nm and $P_y=2000$ nm to preserve the propagation and geometric phases orthogonally.

## 3. Simulation result

To incorporate LP and CP detection from the same metasurface at the different pixel positions of the detector, we finally came to a design as shown in Fig 4(a). LP detector positions D1(ϴ,ϕ) = (26˚, -145˚), D2(ϴ,ϕ) = (26˚, 36˚), CP detector positions D3(ϴ,ϕ) = (26˚, -35˚), D4(ϴ,ϕ) = (26˚, 145˚), as we can see in the far field projection in Fig. 2(a). So, the final design can measure arbitrary LP and CP states simultaneously. The calculated Stokes parameter is shown in Fig 4(b) and Fig 4(c), where we use the following Stokes parameter calculation. The Jones vector representation of the E field components [33,39]:

$$E = \begin{bmatrix} E_x \\ E_y \end{bmatrix} = \begin{bmatrix} E_{0x}e^{-i\varphi_x} \\ E_{0y}e^{-i\varphi_y} \end{bmatrix} \quad (5)$$

Where $E_{0x}$, $E_{0y}$, $\varphi_x$ and $\varphi_y$ are the orthogonal amplitudes and phases respectively. When the phase difference between the $E_x$ and $E_y$ is δ, we can express the fraction of polarized and unpolarized light as the state of the polarization (SOP) by the matrix form for the four Stokes parameters. [4,39–42]:

$$S = \begin{bmatrix} S_0 \\ S_1 \\ S_2 \\ S_3 \end{bmatrix} = \begin{bmatrix} E_x^2 + E_y^2 \\ E_x^2 - E_y^2 \\ 2E_xE_y\cos\delta \\ 2E_xE_y\sin\delta \end{bmatrix} = \begin{bmatrix} I_x + I_y \\ I_x - I_y \\ I_{L_{45}} - I_{L_{-45}} \\ I_{RCP} - I_{LCP} \end{bmatrix} \quad (6)$$

Unlike the randomly distributed unpolarized scalar light field, the polarized light field is a vector [40]. In optical frequency, we can get the measurable parameter intensity only. $S_0$ is the total intensity, $S_1$ indicates the difference in the intensity component $I_x$ and $I_y$, $S_2$ indicates the difference in the intensity component $I_{Linear45^0}$ and $I_{Linear135^0}$, $S_3$ indicates the difference in the intensity of component $I_{RCP}$ and $I_{LCP}$. We can see all these intensities on the different order spot from the far field view in FDTD simulation from this Ge metasurface of this schematic in Fig. 4(a). In the detector plane measuring the intensity of the light on the four different diffraction order pixel positions for TM, TE, LCP, and RCP incidence separately, we can find Stoke parameter Fig. 4(c) in the middle one. As a Ge-grown technique for achieving a homogeneous refractive index for thick film is still under investigation and almost no n, k measured value reported with the experimental measurement after Ge deposition at 4 µm wavelength, we check the tolerance of $n_{Ge}$ from 3.9 to 4.1 to see if the Stokes parameter remains in the range. This design works fine for this $n_{Ge}$ range, as shown in Fig. 4(c). It also has a height tolerance from 2100-2200 nm with both $CaF_2$ and $Al_2O_3$ substrate for the 4 µm working wavelength, which we checked by the Stokes parameter. The Stokes parameters are basically the output intensity contrasts for the orthogonal polarization states. For convenience, the transmission for design update from LP separation and LP+CP separation for both LP and CP incidence is also shown in Fig. 4(b).

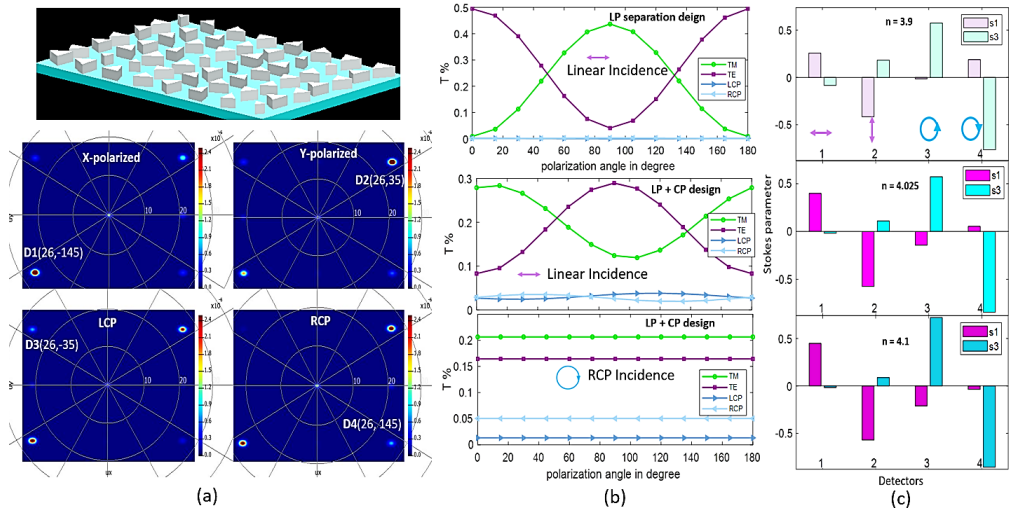

Fig. 4. (a) Schematic of the metasurface and the far-field angular projection for both LP and CP incidence. (b) Transmission efficiency from the designated order position for polarization rotation with LP and CP incidence. (c) calculated Stokes parameter from the simulation result.

The role of the polarization-sensitive camera is to capture the stoke vector from each point of an image. So, all independent polarization state imaging is required to form the stoke vector. From the sensing perspective, the dedicated pixel should record the intensity of the independent polarization state. [4,41]. We want our metasurface to separate the projected most prominent transmission order, based on the polarization states of the incident light, to different pixel positions on the sensor for the intensity vector:

$$I = [I_0\ I_1\ I_2\ I_3]^T \qquad (7)$$

We directly reconstruct the Stokes vector from the simulation by equation (6). In the experimental situation, there is a linear correlation between the Stokes vector and the intensity with instrument matrix A, which comes from the experimental setup for calibration of the polarization states of the incident beam. Knowing the calibration of pure polarization states with our metasurface, we can calculate the Stokes vector by:

$$S = A^{-1}I \qquad (8)$$

After knowing the S values at each pixel, we can find the necessary parameter for the polarization imaging; one is the degree of polarization (DOP), and the other is the physical orientation angle ψ of the polarization ellipse:

$$DOP = \frac{(S_0^2 + S_1^2 + S_2^2 + S_3^2)^{\frac{1}{2}}}{S_0} \qquad (9)$$

$$\psi = \frac{1}{2} tan^{-1}\left(\frac{S_2}{S_1}\right) \qquad (10)$$

The image formed with the total intensity $S_0$ is a regular camera image. However, the image formed with the information of ψ and DOP can reveal more contrast, edge features, and internal stress as that counts the orientation of the electric field in the Poincaré space. We want to do the division of focal plane imaging, for which the spatial phase profile of this type of metasurface of the wavelength λ can be given by [34,43–46].

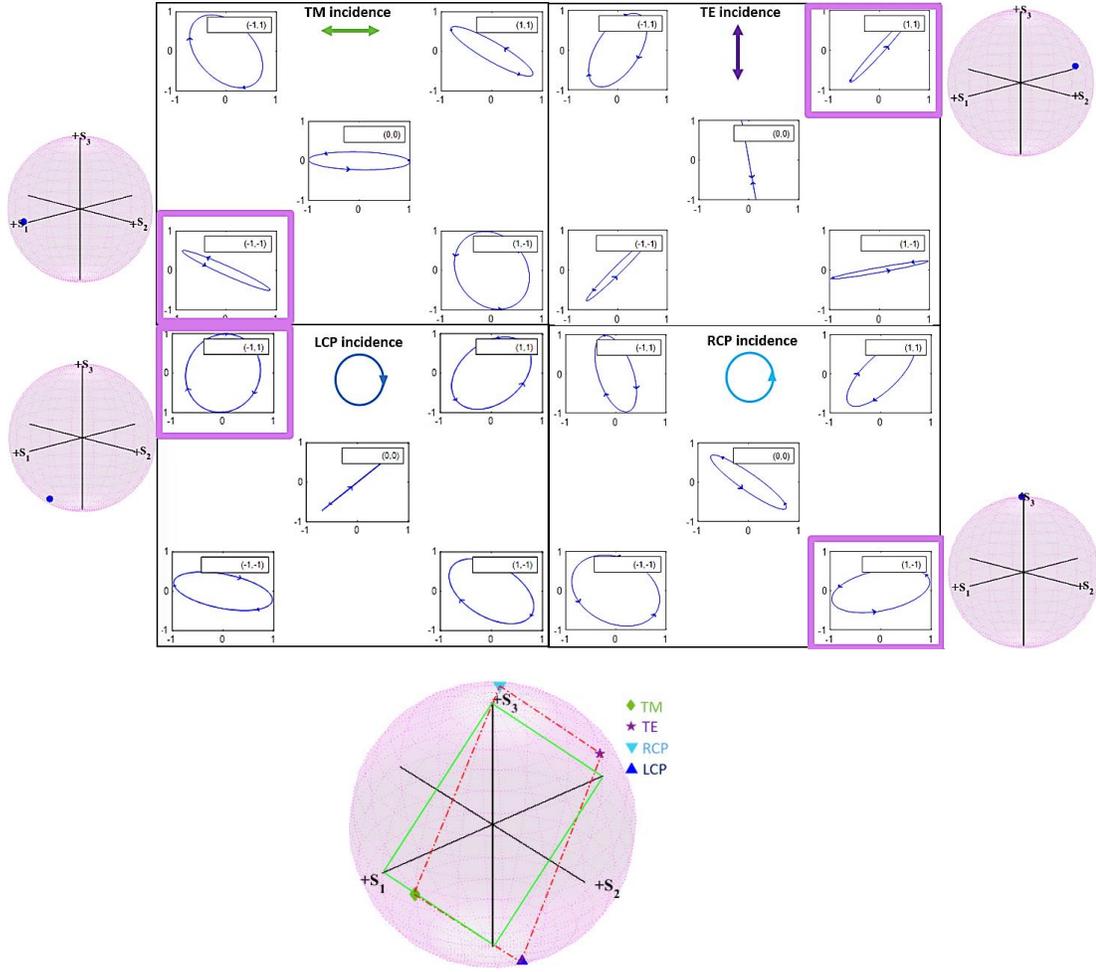

Fig. 5. Simultaneous polarization ellipse at the four designated polarization order positions for a particular incidence polarization. Prominent order for TM at (-1,-1), TE at (1,1), RCP at (1,-1), LCP at (-1,1) position, and all (0,0) order position. Outside Poincaré spheres are the reconstructed Stokes vector for those prominent orders outlined as pink. The bottom Poincaré sphere is the metasurface-analyzed Stokes vector from all the order positions. The green square is the maximally different polarization states, and the dotted red line is the polarimetric reconstruction of the metasurface analyzer polarization states based on the simulated intensity.

In the detector plane measuring the intensity of the light on the different diffraction order pixel positions, we can reconstruct the Stoke parameter, which has been presented from a far-field view in simulation in Fig. 5. Here, the four boxes represent the shape of the output polarization ellipse when one of the four-incidence polarization projected by the metasurface at the detector plane. We also show the (0,0) order position for convenience, which gets negligible intensity for our metasurface. The pink boxed order position is the prominent order reserved to analyze the Stokes parameter. Outside the box, we see the reconstructed Stokes vector in the Poincaré sphere for the designated order position. Although this metasurface does not work for all six input polarization separations because the efficiency and intensity contrast are very low for L+45 and L-45 states, in the calculation, we consider all the intensities from the six polarization states. Finally, the bottom Poincaré sphere shows the figure of merit of this metasurface, where we use the four maximally different input polarization states and reconstruct the Stokes vector from all the order positions by simulated intensity. The deviation in the metasurface analyzer polarization states by the simulated intensity could be due to the loss of the intensity to different unwanted order states apart from the four prominent orders.

For non-paraxial incidence, the performance remains almost the same for 5˚ tilted incident angles, with less than 10% transmission deviation, as presented in Fig. 6. Still, the angular position of the beam changes by about 7˚ as we can see in Fig. 6 far-field projection. For a 10˚ tilted incident angle, the angular position changes to about 14.5˚, so there is an overall 14˚ angular shift for ±5˚ tilted beam and a 29˚ angular shift for ±10˚ tilted beam. From the left side, we see the variation of the Stokes parameter is within tolerable range for the 4-detector position within ±5˚ non-paraxial incidence for this metasurface.

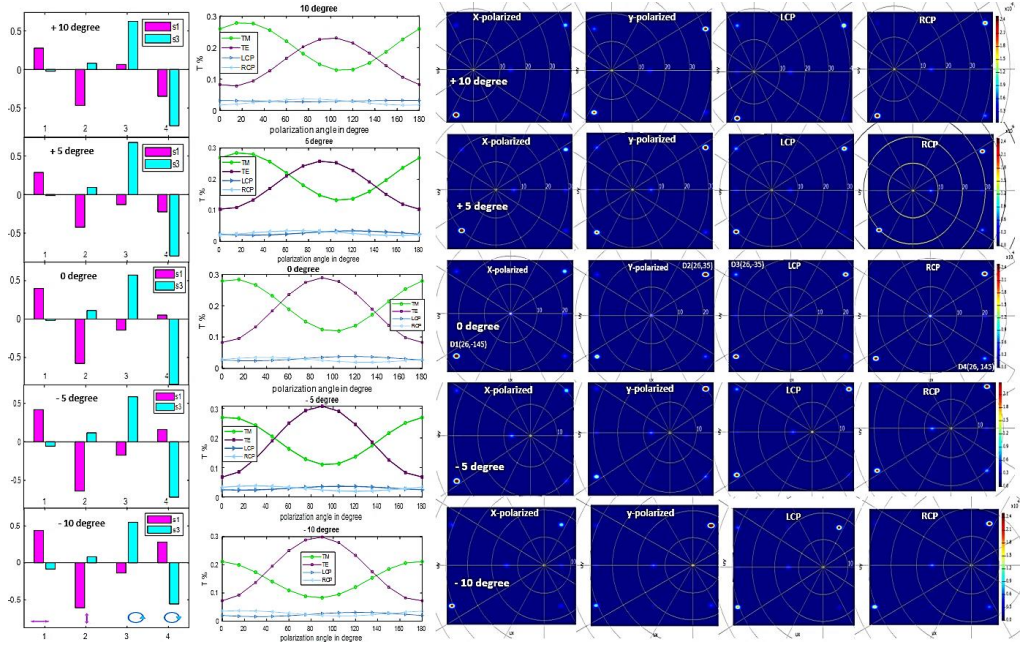

Fig. 6. Stokes parameter (left), transmission (middle), and far-field angular projection (right) for from 10˚ (top) to -10˚ (bottom) tilted incident polarization with 0-180˚ polarization rotation. The designated polarization states to detect are indicated underneath as the detector number for the left picture,

## 4. Conclusion and discussion

Using an asymmetric Ge meta-element, we optimized the polarimetric metasurface design for 4 μm wavelength by significantly reducing the aspect ratio for a structurally robust design for a single-step E-beam lithography fabrication technique, which will provide higher transmission while overcoming the challenges of insufficient phase control. Along the way, we investigate the effect of lattice resonance in terms of periodicity for the interactive meta-element for high-contrast dielectric phase modulation metasurface. Adjusting the propagation and geometric phase orthogonally, we prove that the single-unit cell metasurface works for polarimetric detection by reconstructing the full Stoke parameter from the simulation. Hence, they work simultaneously to detect and analyze any arbitrary polarization state without a shared aperture. The optimized Stokes parameter for refractive index and height tolerances with accommodating slightly angled non-paraxial rays reported a feasible design which, when integrated with the uncooled FPA, can be a chip-scale low-cost for polarization-sensitive thermal imaging device, which is still not available commercially. This technology will open the door for MWIR long-range surveillance, land mine detection, infrared homing devices, and sensors with a compact on-chip polarimetric detection technology.

## 5. Back matter

**Funding:** This research received no external funding.

**Disclosures.** The author declares no competing financial interest.

**Data availability:** No data was generated or analyzed in the presented research.

**Supplemental document.** N/A

## 6. References:

1. F. F. Carvalho, C. A. de M. Cruz, G. C. Marques, and K. M. C. Damasceno, "Angular light, polarization and stokes parameters information in a hybrid image sensor with division of focal plane," Sensors (Switzerland) **20**(12), 1–19 (2020).

2. B. Cheng, Y. Xu, and G. Song, "High spatial resolution Stokes metasurface based on three-pixel technology," Opt Mater Express **13**(5), 1189–1200 (2023).

3. J. L. Pezzaniti and D. B. Chenault, "A division of aperture MWIR imaging polarimeter," Polarization Science and Remote Sensing II **5888**(August 2005), 58880V (2005).

4. N. A. Rubin, G. D'Aversa, P. Chevalier, Z. Shi, W. T. Chen, and F. Capasso, "Matrix Fourier optics enables a compact full-Stokes polarization camera," Science (1979) **364**(6448), (2019).

5. I. Karakasoglu, M. Xiao, and S. Fan, "Polarization control with dielectric helix metasurfaces and arrays," Opt Express **26**(17), 21664 (2018).

6. B. Yang, W. Liu, Z. Li, H. Cheng, S. Chen, and J. Tian, "Polarization-Sensitive Structural Colors with Hue-and-Saturation Tuning Based on All-Dielectric Nanopixels," Adv Opt Mater **6**(4), 1–8 (2018).

7. L. Zhang, X. Q. Chen, S. Liu, Q. Zhang, J. Zhao, J. Y. Dai, G. D. Bai, X. Wan, Q. Cheng, G. Castaldi, V. Galdi, and T. J. Cui, "Space-time-coding digital metasurfaces," Nat Commun **9**(1), 1–11 (2018).

8. Y. Fu, C. Min, J. Yu, Z. Xie, G. Si, X. Wang, Y. Zhang, T. Lei, J. Lin, D. Wang, H. P. Urbach, and X. Yuan, "Measuring phase and polarization singularities of light using spin-multiplexing metasurfaces," Nanoscale **11**(39), 18303–18310 (2019).

9. W. Cao, X. Yang, and J. Gao, "Broadband polarization conversion with anisotropic plasmonic metasurfaces," Sci Rep **7**(1), (2017).

10. K. Zhang, Y. Wang, Y. Yuan, and S. N. Burokur, "A Review of orbital angular momentum vortex beams generation: From traditional methods to metasurfaces," Applied Sciences (Switzerland) **10**(3), (2020).

11. A. C. Overvig, S. Shrestha, S. C. Malek, M. Lu, A. Stein, C. Zheng, and N. Yu, "Dielectric metasurfaces for complete and independent control of the optical amplitude and phase," Light Sci Appl **8**(1), 92 (2019).

12. K. Ou, F. Yu, G. Li, W. Wang, A. E. Miroshnichenko, L. Huang, P. Wang, T. Li, Z. Li, X. Chen, and W. Lu, "Mid-infrared polarization-controlled broadband achromatic metadevice.," Sci Adv **6**(37), (2020).

13. L. Tong, X. Huang, P. Wang, L. Ye, M. Peng, L. An, Q. Sun, Y. Zhang, G. Yang, Z. Li, F. Zhong, F. Wang, Y. Wang, M. Motlag, W. Wu, G. J. Cheng, and W. Hu, "Stable mid-infrared polarization imaging based on quasi-2D tellurium at room temperature," Nat Commun **11**(1), 2308 (2020).

14. H. Zou and G. R. Nash, "Efficient mid-infrared linear-to-circular polarization conversion using a nanorod-based metasurface," Opt. Mater. Express **12**(12), 4565–4573 (2022).

15. M. Dai, C. Wang, B. Qiang, F. Wang, M. Ye, S. Han, Y. Luo, and Q. J. Wang, "On-chip mid-infrared photothermoelectric detectors for full-Stokes detection," Nat Commun **13**(1), (2022).

16. C. Yan, X. Li, M. Pu, X. Ma, F. Zhang, P. Gao, K. Liu, and X. Luo, "Midinfrared real-time polarization imaging with all-dielectric metasurfaces," Appl Phys Lett **114**(16), (2019).

17. J. Chen, F. Yu, X. Liu, Y. Bao, R. Chen, Z. Zhao, J. Wang, X. Wang, W. Liu, Y. Shi, C.-W. Qiu, X. Chen, W. Lu, and G. Li, "Polychromatic full-polarization control in mid-infrared light," Light Sci Appl **12**(1), 105 (2023).

18. Z. Yin, F. Chen, L. Zhu, K. Guo, F. Shen, Q. Zhou, and Z. Guo, "High-efficiency dielectric metasurfaces for simultaneously engineering polarization and wavefront," J Mater Chem C Mater **6**(24), 6354–6359 (2018).

19. K. Guo, X. Li, H. Ai, X. Ding, L. Wang, W. Wang, and Z. Guo, "Tunable oriented mid-infrared wave based on metasurface with phase change material of GST," Results Phys **34**, (2022).

20. K. Ou, F. Yu, G. Li, W. Wang, A. E. Miroshnichenko, L. Huang, P. Wang, T. Li, Z. Li, X. Chen, and W. Lu, "Mid-infrared polarization-controlled broadband achromatic metadevice.," Sci Adv **6**(37), (2020).

21. M. I. G. Arcia, C. H. E. Dmiston, R. A. M. Arinov, A. L. V Ail, and V. I. G. Ruev, "Bio-inspired color-polarization imager for real-time in situ imaging," **4**(10), 15–18 (2017).

22. H. Sultana, "Periodicity and Lattice Resonance: Transmission Control of the High Contrast Dielectric Metasurface," in *Frontiers in Optics + Laser Science 2022 (FIO, LS)* (Optica Publishing Group, 2022), p. JW4B.69.

23. S. Shen, Z. Ruan, S. Li, Y. Yuan, and H. Tan, "The influence of periodicity on the optical response of cube silicon metasurfaces," Results Phys **23**, 104057 (2021).

24. A. B. Evlyukhin and V. E. Babicheva, "Resonant Lattice Kerker Effect in Metasurfaces With Electric and Magnetic Optical Responses," Laser Photon Rev **11**, (2017).

25. M. Kerker, D.-S. Wang, and C. L. Giles, "Electromagnetic scattering by magnetic spheres," J Opt Soc Am **73**(6), 765–767 (1983).

26. Y. H. Fu, A. I. Kuznetsov, A. E. Miroshnichenko, Y. F. Yu, and B. Luk'yanchuk, "Directional visible light scattering by silicon nanoparticles," Nat Commun **4**(1), 1527 (2013).

27. A. B. Evlyukhin, C. Reinhardt, A. Seidel, B. S. Luk'yanchuk, and B. N. Chichkov, "Optical response features of Si-nanoparticle arrays," Phys Rev B **82**(4), 45404 (2010).

28. G. W. Castellanos, P. Bai, and J. Gómez Rivas, "Lattice resonances in dielectric metasurfaces," J Appl Phys **125**(21), 213105 (2019).

29. V. E. Babicheva and J. V Moloney, "Lattice effect influence on the electric and magnetic dipole resonance overlap in a disk array," **7**(10), 1663–1668 (2018).

30. S. Tsoi, F. J. Bezares, A. Giles, J. P. Long, O. J. Glembocki, J. D. Caldwell, and J. Owrutsky, "Experimental demonstration of the optical lattice resonance in arrays of Si nanoresonators," Appl Phys Lett **108**(11), 111101 (2016).

31. N. Yu, P. Genevet, M. a Kats, F. Aieta, J.-P. Tetienne, F. Capasso, and Z. Gaburro, "Light Propagation with Phase Reflection and Refraction," Science (1979) **334**(October), 333–337 (2011).

32. H. Sultana, "Coupled Plasmon Wave Dynamics beyond Anomalous Reflection: A Phase Gradient Copper Metasurface for the Visible to Near-Infrared Spectrum," Optics **3**(3), 243–253 (2022).

33. F. Ding, S. Tang, and S. I. Bozhevolnyi, "Recent Advances in Polarization-Encoded Optical Metasurfaces," Adv Photonics Res **2**(6), 2000173 (2021).

34. J. P. Balthasar Mueller, N. A. Rubin, R. C. Devlin, B. Groever, and F. Capasso, "Metasurface Polarization Optics: Independent Phase Control of Arbitrary Orthogonal States of Polarization," Phys Rev Lett **118**(11), 1–12 (2017).

35. D. Wen, F. Yue, S. Kumar, Y. Ma, M. Chen, X. Ren, P. E. Kremer, B. D. Gerardot, M. R. Taghizadeh, G. S. Buller, and X. Chen, "Metasurface for characterization of the polarization state of light," Opt Express **23**(8), 10272–10281 (2015).

36. M. Kang, T. Feng, H.-T. Wang, and J. Li, "Wave front engineering from an array of thin aperture antennas," Opt Express **20**(14), 15882–15890 (2012).

37. Y.-C. Chen, Q.-C. Zeng, C.-Y. Yu, and C.-M. Wang, "General case of the overall phase modulation through a dielectric PB-phase metasurface," OSA Contin **4**(12), 3204 (2021).

38. H. Sultana and B. Weng, "Implementing the Geometric Phase for Designing the Axially Asymmetric Metasurface Element," in *CLEO 2024*, Technical Digest Series (Optica Publishing Group, 2024), p. JTu2A.164.

39. E. Arbabi, S. M. Kamali, A. Arbabi, and A. Faraon, "Full-Stokes Imaging Polarimetry Using Dielectric Metasurfaces," ACS Photonics **5**(8), 3132–3140 (2018).

40. H. Duan, Y. Hu, X. Wang, X. Luo, X. Ou, L. Li, Y. Chen, P. Yang, and S. Wang, "All-dielectric metasurfaces for polarization manipulation: Principles and emerging applications," Nanophotonics **9**(12), 3755–3780 (2020).

41. S. H. W. Ei and Z. H. Y. Ang, "Design of ultracompact polarimeters based on dielectric metasurfaces," **42**(8), 1580–1583 (2017).

42. N. A. Rubin, P. Chevalier, M. Juhl, M. Tamagnone, R. Chipman, and F. Capasso, "Imaging polarimetry through metasurface polarization gratings," Opt Express **30**(6), 9389 (2022).

43. H.-H. Hsiao, C. H. Chu, and D. P. Tsai, "Fundamentals and Applications of Metasurfaces," Small Methods **1**(4), 1600064 (2017).

44. J. Zhang, J. Zeng, Y. Liu, Y. Dong, and J. Wang, "Fundamental challenges induced by phase modulation inaccuracy and optimization guidelines of geometric phase metasurfaces with broken rotation symmetry," Opt Express **29**(21), 34314 (2021).

45. P. Lalanne and P. Chavel, "Metalenses at visible wavelengths: past, present, perspectives," Laser Photon Rev **11**(3), (2017).

46. F. Aieta, P. Genevet, M. A. Kats, N. Yu, R. Blanchard, Z. Gaburro, and F. Capasso, "Aberration-free ultrathin flat lenses and axicons at telecom wavelengths based on plasmonic metasurfaces," Nano Lett **12**(9), 4932–4936 (2012).